\documentclass[preprint,authoryear,12pt]{elsarticle}

\usepackage{amssymb}
\usepackage{amssymb}
\usepackage{latexsym}
\usepackage{amsmath}
\usepackage{graphicx}

\begin{document}

\begin{frontmatter}



\title{Anti-plane Shear Waves in Layered Composites:\\
Band Structure and Anomalous  Wave-refraction}



\author{Sia Nemat-Nasser}

\address{Department of Mechanical and Aerospace Engineering\\ 
University of California, San Diego\\
La Jolla, CA, 92093-0416 USA}

\date{\today}

\begin{abstract}

For oblique anti-plane shear waves in periodic layered elastic composites, it is shown that \textit{negative energy refraction is  accompanied by positive phase-velocity refraction} and
\textit{positive energy refraction is  accompanied by negative phase-velocity refraction}, and that, this happens over a broad range of frequencies. 
The composite's unit cell may consist of any number of layers of any variable  mass-density and elastic shear modulus (with large discontinuities). 
  
Explicit series expressions for displacement, velocity, strain and stress components,  and energy-flux fields are given, and group-velocity vector is calculated. 
The approach is based on a mixed variational principle where 
the displacement and stress components are viewed as independent fields subject to arbitrary variation. These fields are hence approximated independently, thereby ensuring the necessary continuity conditions. 
The resulting computational method yields the composite's frequency band structure and the associated mode shapes, in terms of the wave-vector components for any desired number of frequency bands.  The calculations are direct and require no iteration, accurately and efficiently producing the entire band structure of the composite.  
This also allows for direct calculation of the components of the group-velocity, energy-flux, and phase-velocity vectors as functions of the frequency and wave-vector components, over an entire frequency band. 

The general results are illustrated using a two-phase and a three-phase unit cell with piecewise constant properties. It is shown that the directions of the group-velocity and energy-flux vectors are essentially indistinguishable for this class of problems, and, 
more importantly that, 
on their second frequency pass-bands, 
\textit{only the components of the phase and group velocities normal to the layers are antiparallel, while the components along the layers are parallel.} 
Therefore,
both the two-phase and the three-phase composites \textit{display negative energy refraction with positive phase-velocity refraction} and \textit{positive  phase-velocity refraction with negative energy refraction},
depending on how the composite is interfaced with a homogeneous solid.

The presented method is applicable and effective also when some or all of the layers in a unit cell have spatially varying properties. 


\end{abstract}


\end{frontmatter}



\newpage

\section{Introduction}\label{Introduction}

Elastic composites can have remarkable mechanical and acoustic properties that are not shared by their individual constituents.  
They have found widespread applications and hence have been extensively studied; see for example (\cite{willis1981variationala}, \cite{christensen2012mechanics}, \cite{nemat1999micromechanics}, \cite{maldovan2009periodic}, \cite{banerjee2011introduction}, \cite{green1991reflection}). 
Here the focus is on harmonic waves in layered elastic composites, and their dynamic  properties (\cite{willis1981variationalb}, \cite{nayfeh1995wave}, \cite{milton2007modifications}, \cite{willis2013study}). 
In this context, efficient and accurate 
calculation of their band structure  for Bloch-form harmonic waves is the 
necessary first step in estimating their overall response and effective dynamic constitutive properties (\cite{nemat2011overall}, \cite{PRBnemat2011}).  
Various techniques have been used to study the dynamic responses of laminated composites; see \cite{sigalas2005classical} for a review of recent contributions.  
This includes, for example, methods such as 
direct analytic solutions for layered composites (\cite{rytov1956acoustical}, \cite{mal1988wave}, \cite{braga1992floquet});
transfer matrix (\cite{thomson1950transmission}, \cite{gilbert1966propagator}, \cite{bahar1972transfer}, \cite{hosten1993transfer}, \cite{rokhlin2002stable}); 
plane-wave expansion (\cite{nayfeh1991general}); 
displacement-based variational methods (\cite{goffaux2003two}, \cite{goffaux2004comparison});  
and 
finite elements (\cite{langlet1995analysis}, \cite{aaberg1997usage}, \cite{aboudi1986harmonic}).  

A general feature of layered composites is the presence of finite discontinuities in the properties (mass-density and stiffness) of their constituents.  This renders the application of the usual Rayleigh quotient computationally ineffective (\cite{kohn-krumhansl-lee1972}, \cite{SNNdiscussion1973}).

To produce an effective tool that accounts for discontinuities as an integral part of the variational formulation, a mixed variational method has been developed and successfully used in the 1970's to calculate the band structure of one, two, and three-dimensional periodic elastic composites;  (\cite{NN-1972a}, \cite{nemat1972harmonic}, \cite{ nemat1975harmonic},  \cite{minagawa1976harmonic}). 
This  mixed variational method yields very accurate results and the rate of convergence of the corresponding approximating series solution is greater than that of  the Rayleigh quotient with displacement-based approximating functions (\cite{babuvska1978numerical}).  
Since it is based on a variational principle, any set of approximating functions can be used for calculations, e.g., plane-waves Fourier series, as in the above cited papers, or finite elements (\cite{minagawa1981finite}).

The method has been revived in recent years and applied to calculate the effective overall dynamic constitutive parameters of  periodic composites for Bloch waves traveling normal to the layers (\cite{PRBnemat2011}). The limits of the accuracy of the resulting estimates have been examined by  considering the response of a layered composite  interfaced with its homogenized half-space (\cite{srivastava2014limits}; see also, \cite{willis2013some}).  

When elastic waves are at an angle relative to an interface of a half-space layered composite,
they generate a complex set of reflected and transmitted waves due to the inherent structure of the layered (or its homogenized) medium. 
It was recently suggested by Willis (2013b) that this complexity is avoided by considering oblique anti-plane shear waves.    This then allows the  study of a number of physically interesting phenomena, such as negative refraction, within a relatively simple mathematical framework.  Willis (2013b) shows negative energy refraction when a layered composite is interfaced with a homogeneous solid on a plane normal to the layers. Here, we revisit this and in addition show that, unlike metamaterials, such negative refraction is accompanied by positive phase-velocity refraction.

The calculation of the  band structure, mode shapes, group-velocity, and energy-flux vectors
in terms of the wave-vector components for a wide range of frequencies, is  a challenging task.
Here this problem is addressed,  successfully formulated, and solved using a mixed variational method to calculate \textit{the entire band structure and the associated mode shapes} for oblique anti-plane shear waves in layered elastic composites. The  composite may consist of periodically distributed unit cells of any number of layers of any desired properties that may vary in the direction normal to the layers.  The results are then used to study the overall dynamic response of this class of composites.   
The problem is formulated for general unit cells and the results are illustrated  using two-phase and three-phase unit cells. It is shown by direct calculation that the method easily produces any desired frequency band in terms of the wave-vector components and/or the angle that the wave vector makes with the direction normal to the layers. For illustration, numerical results for two- and three-phase unit cells are worked out in detail, where the material properties of each layer are uniform. 

It is shown that composites with \textit{two- and three-phase unit cells may display negative refraction}, which however is accompanied by \textit{positive phase-velocity  refraction}.  Indeed, for this class of composites, it is shown that \textit{only the components of the phase and group velocities normal to the layers are antiparallel while the corresponding components along the layers are parallel.}   
This phenomenon is demonstrated by 
considering the refraction and reflection of plane waves when the composite is in contact with a homogeneous solid on a plane normal to the layers, as well as when the contact plane is parallel to the layers. In the first case, negative energy refraction with positive phase refraction may occur, and, in the second case, positive energy refraction with negative phase refraction may occur. 

The calculations are direct and require no iteration, producing the entire band structure of the composites with unit cells of any number of layers of any constant or variable properties.  This also 
yields explicit series-form expressions for all the field variables necessary to 
calculate  the components of the group-velocity and the energy-flux vectors, from which the overall response of the composite can readily be extracted. 

When the phase-velocity  and energy-flux vectors are antiparallel, the resulting waves have been called \textit{backward waves} or BW (\cite{oliner1962backward}, \cite{lindell2001bw}). 
In general, backward waves and negative refraction occur when 
the wave-vector and the energy-flux vector are antiparallel.
Here, for the layered composite, we show 
\textit{negative energy  refraction} with
\textit{positive  phase refraction}, that is, we show
that only one component of the phase velocity is antiparallel with its corresponding component of the energy-flux (or group velocity) vector while the other component of these vectors are parallel, leading to negative refraction with positive  phase  refraction.  A similar phenomenon was first recognized to exist in photonic crystals by \cite{meisels2005refraction}. 
In addition, we show here that \textit{negative phase refraction} can occur with \textit{positive energy refracttion}.

\section{Statement of the Problem and Field Equations}

Consider a layered composite and take the $x_1$-axis normal, and the $x_2$ and $x_3$ parallel to the layers. With $a$ denoting the length of a typical unit cell, the mass-density $\hat{\rho}$ and the elastic shear moduli, $\hat{\mu}_{jk}$, $j, k = 1, 2$, with $\hat{\mu}_{12} =\hat{ \mu}_{21}$, have the periodicity of the composite, i.e.,
\begin{equation}
\hat{\rho}(x_1)=\hat{\rho}(x_1+ma);\quad\hat{\mu}_{jk}(x_1)=\hat{\mu}_{jk}(x_1+ma), \quad j,k=1,2,
\end{equation}
for any integer m.

For Bloch-form time-harmonic anti-plane shear waves of frequency $\omega$ and wave-vector  components $k_1$ and $k_2$, the nonzero displacement component  $\hat{u}_3 (x_1,x_2,t)$ has the following structure:
\begin{equation}\label{HatW}
\hat{u}_3=u_3^{p}(x_1)e^{i(k_1x_1+k_2x_2-\omega t)},
\end{equation}
where $u_3^{p}(x_1)$ is periodic with the periodicity of the unit cell. 

Set 
\begin{equation}\label{Dimensionless}
\xi_1=x_1/a,\quad\xi_2=x_2/a,\quad Q_1=k_1a\quad Q_2=k_2a,
\end{equation}
introduce the average parameters,
\begin{equation}\label{Average-1}
{\bar{\rho}}=\int_{-1/2}^{1/2}\hat{\rho}(a\xi)d\xi,\quad\
\bar{\mu}_{11}=\int_{-1/2}^{1/2}\hat{\mu}_{11}(a\xi)d\xi,
\end{equation}
and consider the following dimensionless quantities:
\begin{equation}\label{DimensionLess-1}
\begin{split}
{\mu}_{jk}(\xi_1)=\hat{\mu}_{jk}(a\xi_1)/\bar{\mu}_{11},\quad
\rho(\xi_1)=\hat{\rho}(a\xi_1)/\bar\rho,\qquad
\nu^2=a^2\omega^2\bar{\rho}/\bar{\mu}_{11},\quad\\
\begin{aligned}
w^{p}(\xi_1)=u_3^{p}(a\xi_1)\sqrt{\bar{\mu}_{11}},\qquad
\tau_1=\sigma_{13}/\sqrt{\bar{\mu}_{11}},\qquad 
\tau_2=\sigma_{23}/\sqrt{\bar{\mu}_{11}},\qquad
\end{aligned}
\end{split}
\end{equation}
where $\nu$ is the dimensionless frequency.  Here the displacement, $u_3^{p}$, and the nonzero (shear) stresses, $\sigma_{13}=\sigma_{31}$ and $\sigma_{23}=\sigma_{32}$, are rendered nondimensional and denoted by $w$ and $\tau_j$, $j=1,2$, respectively. 
In what follows, the corresponding (engineering) strains, $2\epsilon_{13}=2\epsilon_{31}$ and $2\epsilon_{23}=2\epsilon_{32} $,  will be denoted by $\gamma_1$ and $\gamma_2$, respectively.
The (normalized) field equations then become,
\begin{equation}\label{EFieldEqn1}
\tau_{1,1}+\tau_{2,2}+\nu^2\rho w=0;\quad 
\tau_{j}=\mu_{jk} \gamma_k\quad
(k \ summed),
\end{equation}
\begin{equation}\label{EFieldEqn2}
\gamma_1 = { w}_{,1}\quad 
\gamma_2=iQ_2w,\quad 
w=w^pe^{iQ_1\xi_1}.
\end{equation}

In view of equations (\ref{EFieldEqn1}, \ref{EFieldEqn2}) and (\ref{HatW}), the shear stress $\tau_2$ can be expressed in terms of $\tau_1$ and $ w$, as follows:
\begin{equation}\label{Tau2}
\tau_2=\frac{\mu_{12}}{\mu_{11}}\tau_1+\frac{iQ_2}{D_{22}}w,
\end{equation}
where $D_{22}$ is the corresponding component of the normalized elastic compliance matrix defined by,
\begin{eqnarray*}\label{Compliance}
\left(
\begin{array}{cc}
  D_{11}  \ D_{12} \\
  D_{21}  \ D_{22}   \\   
\end{array}
\right)=\frac{1}{\Delta}
\left(
\begin{array}{cc}
\mu_{22}  \ \ -\mu_{12} \\
-\mu_{21}  \ \ \mu_{11}   \\  
\end{array}
\right),\quad \Delta=\mu_{11}\mu_{22}-\mu^2_{12}.
\end{eqnarray*}
Hence, 
\begin{equation}\label{tau2,2}
\tau_{2,2}=iQ_2\tau_2=iQ_2\frac{\mu_{12}}{\mu_{11}}\tau_1-\frac{{Q_2}^2}{D_{22}} w.
\end{equation}

\section{Variational Formulation}

Consider now the following functional:
\begin{equation}\label{Variation}
I=<\tau_j,{w}_{,j}>+<{w}_{,j},\tau_j>-<D_{jk} \tau_k,\tau_j>-\nu^2 <\rho {w},{w}>,
\end{equation}
where $<g u,v> = \int_{-1/2}^{1/2} guv^*d\xi$ for a real-valued function $g(\xi)$ and complex-valued functions $u(\xi)$ and $v(\xi)$, with star denoting complex conjugate. 
In (\ref{Variation}) $w$ and $ {\tau}_j $ are viewed as independent fields subject to arbitrary variations.  
It is easy to show (\cite{nemat1975harmonic}) that equations (\ref{EFieldEqn1}) are the Euler equations that render the functional $I$ stationary. 
It is however, expedient to use (\ref{Tau2}) in (\ref{Variation}) and consider $\tau_1$ and ${w}$ as the only independent fields subject to variations, especially since the periodicity in layered composites is only in  one direction. 
Hence consider the functional,
\begin{equation}\label{Variation-1}
\begin{split}
I_1=<\tau_1,{w}_{,1}>+<{w}_{,1},\tau_1>
-iQ_2<\frac{\mu_{12}}{\mu_{11}} \tau_1,{w}>
+iQ_2<\frac{\mu_{12}}{\mu_{11}} {w},\tau_1>\\
\begin{aligned}
+Q^2_2<\frac{1}{D_{22}}{w},{w}>
-<\frac{1}{\mu_{11}}\tau_1,\tau_1>
-\nu^2<\rho {w},{w}>.
\end{aligned}
\end{split}
\end{equation}
The first variation of $I_1$ with respect to ${w}^*$ and $\tau_1^*$ yields, respectively,
\begin{equation}\label{EulerEqs}
\tau_{1,1}+[iQ_2\frac{\mu_{12}}{\mu_{11}} \tau_1
-\frac{Q^2_2}{D_{22}}{w}]+\nu^2\rho w=0\qquad
\tau_1=\mu_{11}w_{,1}+iQ_2\mu_{12} w,
\end{equation} 
which also follow from (\ref{EFieldEqn1}, \ref{EFieldEqn2}) and (\ref{Tau2}).

To find an approximate solution of the field equations (\ref{EulerEqs}) subject to the Bloch periodicity condition (\ref{HatW}), consider the following estimates:
\begin{equation}\label{Estimates}
{w}=\sum_{\alpha=-M}^{+M}W^{(\alpha)}e^{i(Q_1+2\pi\alpha )\xi_1},
\quad {\tau}_1=\sum_{\alpha=-M}^{+M}T^{(\alpha)}e^{i(Q_1+2\pi\alpha )\xi_1},
\end{equation}
which automatically ensure the Bloch and continuity conditions.  

Substitution into (\ref{Variation-1}) now yields,
\begin{equation}\label{Variation-approx}
\begin{split}
I_1=\sum_{\alpha , \beta =-M}^{+M}
\{
T^{(\alpha)}[-i(Q_1+2\pi \alpha)]\delta_{\alpha \beta} W^{(\beta)*}+
W^{(\alpha)}[+i(Q_1+2\pi \alpha)]\delta_{\alpha \beta} T^{(\beta)*}\\
\begin{aligned}
-iQ_2T^{(\alpha)} \Lambda^{(\alpha \beta)}[\mu_{12}/\mu_{11}] W^{(\beta)*}
+iQ_2W^{(\alpha)} \Lambda^{(\alpha \beta)}[\mu_{12}/\mu_{11}] T^{(\beta)*}\\
-T^{(\alpha)} \Lambda^{(\alpha \beta)}[1/\mu_{11}] T^{(\beta)*}+
W^{(\alpha)}\Lambda^{(\alpha \beta)}[Q^2_2/D_{22}-\nu^2 \rho] W^{(\beta)*}
\},
\end{aligned}
\end{split}
\end{equation}
where $ \Lambda^{(\alpha \beta)}$ is a linear integral operator, defined by
\begin{equation}\label{Operator}
\Lambda^{(\alpha \beta)}[f(\xi)]=
\int_{-1/2}^{1/2}f(\xi) e^{i2\pi (\alpha - \beta)\xi}d\xi =
\int_{-1/2}^{1/2}f(\xi) e^{-i2\pi (\beta- \alpha)\xi}d\xi= 
\Lambda^{(\beta \alpha)^*}[f(\xi)],
\end{equation}
with $f(\xi)$ being a real-valued integrable function.  

For an even function, $f(\xi)=f(-\xi)$ (symmetric unit cells), 
\begin{equation}\label{Symmetric-Op}
\Lambda^{(\alpha \beta)}[f(\xi)]=
2\int_{0}^{1/2}f(\xi) cos(2\pi (\alpha - \beta)\xi)d\xi =
\Lambda^{(\beta \alpha)}[f(\xi)].
\end{equation}
Furthermore, for a piecewise constant $f(\xi)$, e.g.,
\begin{displaymath}
 f(\xi) = \left\{
  \begin{array}{lr}
 	 f_1 & 0<\xi<l_1/2\\
       f_2 &  l_1/2<\xi<l_2/2\\
       ...\\
       ...\\
        f_n &  l_{n-1}/2<\xi<1/2
     \end{array}
   \right.
\end{displaymath}
\begin{equation}\label{Piecewise-1}
\end{equation}
with $l_n=1$, one obtains,
\begin{displaymath}
   \Lambda^{(\alpha \beta)}[f(\xi)]=\left\{
    \begin{array}{lr}
	\sum_{a=1}^{n}\frac{f_a-f_{a+1}}{\pi(\alpha - \beta)}sin(\pi(\alpha -\beta)l_a) & {\alpha} \neq {\beta}\\
	 \sum_{a=1}^{n}(f_a-f_{a+1})l_a=\bar{f}& \alpha = \beta, \quad f_{n+1}=0.
	  \end{array}
   \right.
\end{displaymath}
\begin{equation}\label{SymmetricCell}
\end{equation}
 
Define an $N \times N$ matrix $\boldsymbol{\Lambda}[f(\xi)]=[\Lambda^{(\alpha \beta)}][f(\xi)]$, $\alpha, \beta = 0, \pm {1},...,\pm {M} $,  and $N=2M+1$, and note that, in view of linearity, for any two constants $a_1$ and $a_2$,
$$a_1\boldsymbol{\Lambda}[f_1(\xi)]+a_2\boldsymbol{\Lambda}[f_2(\xi)]=
\boldsymbol{\Lambda}[a_1f_1(\xi)+a_2f_2(\xi)].$$
Also, let $\mathbf{H}$ be an $N\times N$ diagonal matrix with components $(Q_1+2\pi \alpha)\delta_{\alpha \beta}$. Then, (\ref{Variation-approx}) can be rewritten as,
\begin{eqnarray}\label{Variation-3}
I_1=\left[ \begin{array}{c}
\mathbf{W} \\
\mathbf{T}\end{array} \right]^T \
\left[ \begin{array}{cc}
Q_2^2\boldsymbol{\Lambda}[1/D_{22}]-\nu^2\boldsymbol{\Lambda}[\rho]\qquad &
i\{\mathbf{H}+Q_2 \boldsymbol{\Lambda}[\mu_{12}/\mu_{11}]\}\\
-i\{\mathbf{H}+Q_2 \boldsymbol{\Lambda}[\mu_{12}/\mu_{11}]\} &- \boldsymbol{\Lambda}[1/\mu_{11}]\\
\end{array} \right]\
\left[ \begin{array}{c}
\mathbf{W^*} \\
\mathbf{T^*}\end{array} \right].
\end{eqnarray}

For symmetric unit cells, $I_1$ may be written as,
\begin{eqnarray}\label{Variation-Symm}
I_{1s}=\left[ \begin{array}{c}
\mathbf{W^*} \\
\mathbf{T^*}\end{array} \right]^T \
\left[ \begin{array}{cc}
Q_2^2\boldsymbol{\Lambda}[1/D_{22}]-\nu^2\boldsymbol{\Lambda}[\rho]\qquad &
-i\{\mathbf{H}+Q_2 \boldsymbol{\Lambda}[\mu_{12}/\mu_{11}]\}\\
i\{\mathbf{H}+Q_2 \boldsymbol{\Lambda}[\mu_{12}/\mu_{11}]\} &- \boldsymbol{\Lambda}[1/\mu_{11}]\\
\end{array} \right]\
\left[ \begin{array}{c}
\mathbf{W} \\
\mathbf{T}\end{array} \right].
\end{eqnarray}
Furthermore, when $\mu_{12}=0$, we obtain,
\begin{eqnarray}\label{Variation-4}
I_2=\left[ \begin{array}{c}
\mathbf{W^*} \\
\mathbf{T^*}\end{array} \right]^T \
\left[ \begin{array}{cc}
Q_2^2\boldsymbol{\Lambda}[\mu_{22}]-\nu^2\boldsymbol{\Lambda}[\rho]\qquad &
-i\mathbf{H}\\
i\mathbf{H} & -\boldsymbol{\Lambda}[1/\mu_{11}]\\
\end{array} \right]\
\left[ \begin{array}{c}
\mathbf{W} \\
\mathbf{T}\end{array} \right].
\end{eqnarray}
Note that, in view of equation (\ref{Symmetric-Op}), $ \boldsymbol{\Lambda}[f(\xi)]$ in (\ref{Variation-3}) and (\ref{Variation-4}) is real-valued for symmetric unit cells. Furthermore, when a symmetric unit cell consists of layers of uniform elasticities and densities, equation (\ref{SymmetricCell}) gives $\Lambda^{(\alpha \beta)} [f(\xi)] $ explicitly.

Now, minimization of $I_2$ (or $I_1$ for $\mu_{12}\neq 0$) with respect to the unknown coefficients $W^{(\alpha)}$ and $T^{(\alpha)}$, results in an eigenvalue problem which yields the band structure of the composite for anti-plane Bloch-form shear waves,
\begin{equation}\label{Eigen-1}
\left[(\mathbf{I}+Q^2_2\mathbf{A}\boldsymbol{\Lambda}[\mu_{22}])
-\nu^2\mathbf{A}\boldsymbol{\Lambda}[{\rho}]\right]{\mathbf{W}}=\mathbf{0},\qquad
\mathbf{A}=\mathbf{H}^{-1}\boldsymbol{\Lambda}[1/\mu_{11}]\mathbf{H}^{-1},
\end{equation}
where $\mathbf{I}$ is the identity matrix. For given values of $Q_1$ and $Q_2$, the eigenvalues, $\nu$, of equation(\ref{Eigen-1})$_1$ are obtained from 
\begin{equation}\label{Eigen-2}
det\left|(\mathbf{I}+Q^2_2\mathbf{A}\boldsymbol{\Lambda}[\mu_{22}])
-\nu^2\mathbf{A}\boldsymbol{\Lambda}[{\rho}]\right|=0,
\end{equation}
and for each eigenvalue, the corresponding displacement field, ${\mathbf{W}}$, is given by (\ref{Eigen-1})$_1$, and the stress field by
\begin{equation}\label{Talpha} 
{\mathbf{T}}=i\{\boldsymbol{\Lambda}[1/\mu_{11}]\}^{-1}\mathbf{H}{\mathbf{W}}.  
\end{equation}
Note that $\mathbf{H}$ is a diagonal matrix whose components are linear in $Q_1$. 

From (\ref{EFieldEqn1}, \ref{EFieldEqn2})), the $x_2$-component of the shear stress, $\tau_2$, and the $x_1$-component of the shear strain, $\gamma_1$, are given by,
\begin{equation}\label{tau2-gamma_1}
\tau_2=iQ_2\mu_{22}(\xi_1)
\sum_{\alpha=-M}^{+M}W^{(\alpha)}e^{i(Q_1+2\pi\alpha )\xi_1},\quad
\gamma_1=D_{11}(\xi_1) 
\sum_{\alpha=-M}^{+M}T^{(\alpha)}e^{i(Q_1+2\pi\alpha )\xi_1}.
\end{equation}
As can be seen, $\mathbf{W}$ is real-valued and $\mathbf{T}$ is purely imaginary.  Both are implicit functions of $Q_1$ and $Q_2$.

The periodic parts of the displacement, velocity, and stresss-components  
are summarize here for subsequent application,
\begin{equation}\label{w-wdot}
w^p(\xi_{1})=\sum_{\alpha=-M}^{+M}W^{(\alpha)}e^{i2\pi\alpha \xi_1},\quad
\dot{w}^p(\xi_{1})=-i\nu\sum_{\alpha=-M}^{+M}W^{(\alpha)}e^{i2\pi\alpha \xi_1},\quad
\end{equation}
\begin{equation}\label{tau-1_2}
\tau_1^p(\xi_{1})=
\sum_{\alpha=-M}^{+M}T^{(\alpha)}e^{i2\pi\alpha \xi_1},\quad
\tau_2^p(\xi_{1})=iQ_2\sum_{\alpha=-M}^{+M}W^{(\alpha)}\mu_{22}(\xi_1)e^{i2\pi\alpha \xi_1},
\end{equation}
where for each frequency band $J$, associated with an eigenvalue $\nu_J$, equations (\ref{Eigen-1}, \ref{Talpha}) yield the corresponding coefficients, $W_J^{(\alpha)}$ and $T_J^{(\alpha)}$; the subscript $J$ has been omitted in the above expressions.
Once $W^{(\alpha)}$ and $T^{(\alpha)}$ are calculated for a desired eigenvalue, $\nu$, the above expressions give the periodic part of the field variables.

\section{Phase and Group Velocities, and Energy Flux}

For a given (symmetric) unit cell that consists of a given number of layers of prescribed mass densities and stiffnesses, matrices $\mathbf{A}$ and $\boldsymbol{\Lambda}[f(\xi)]$ in (\ref{Eigen-1}) can be computed explicitly using (\ref{Symmetric-Op}). The expression in the right-hand side of (\ref{Eigen-2}) will then depend parametrically on the wave-vector components, $Q_1$ and $Q_2$.  The resulting eigenfrequencies, $\nu$, can thus be expressed as functions of $Q_1$ and $Q_2$. These eigenfrequencies form surfaces in the ($Q_1$, $Q_2$,  $\nu$)-space, referred to as Brillouin zones. The first zone corresponds to $-\pi{\leq{Q_1,Q_2}\leq }\pi$. We focus on this zone and examine the dynamic properties of layered elastic composites on the first and second frequency bands.

On each frequency band, the phase and group velocities are given by,
\begin{equation}\label{ph-group}
v^p_{Jk}=\frac{\nu_JQ_k}{Q_1^2+Q_2^2}  \quad
v^g_{Jk}=\frac{\partial \nu_J}{\partial{Q_k}},\quad
k=1,2;
\end{equation}
here and below, $J=1, 2,...$ denotes the frequency band and $k=1,2$, the $x_1$-  and the $x_2$-directions, respectively.  
The group velocity defines the direction of the energy flux; it is given by given by
\begin{equation}\label{GV-angle}
\alpha_{J}=atan(\frac{v^g_{J2}}{v^g_{J1}}).
\end{equation}

The $x_1$- and $x_2$-components of the energy-flux vector are given by
\begin{equation}\label{Energy-flux}
E_{Jk}=\frac{\nu_J}{2\pi}\int_{0}^{\frac{2\pi}{\nu_J}}
<Re({\tau_{Jk}(\xi_1,\xi_2,t)})Re({\dot{w}_J^*(\xi_1,\xi_2,t)})>dt
=-\frac{1}{2}<\tau_{Jk}^p\dot{w}_J^{p*}>,
\end{equation}
which is real-valued, where $k=1,2$.  Substitution from (\ref{w-wdot}, \ref{tau-1_2}) results in,  
\begin{equation}\label{EF-1_2}
E_{J1}=-\frac{1}{2}i\nu_J \sum_{\alpha=-M}^{+M}T_J^{(\alpha)}W_J^{\alpha},\quad
E_{J2}=\frac{1}{2}\nu_J  Q_2\sum_{\alpha,\beta=-M}^{+M}W_J^{(\alpha)}W_J^{(\beta)}\Lambda^{(\alpha \beta)}[\mu_{22}].
\end{equation}
The direction, $ {\beta_J} $,  of the energy-flux vector is hence given by,
\begin{equation}\label{Refraction_Energy}
\beta_J=atan(\frac{E_{J2}}{E_{J1}}).
\end{equation}
It is known (\cite{brillouin1948wave}) that the direction, $\alpha_J$, is essentially the same as the direction $\beta_J$ of the energy flux for nondissipative media.  We shall illustrate this in what follows.

\textbf{An Important Cautionary Note:}  For an oblique anti-plane shear wave in a periodic layered elastic composite, the angle of incidence $\theta=atan(\frac{Q_2}{Q_1})$ cannot be arbitrary, limiting the admissible values of $Q_2$ depending on the structure and composition of the corresponding  unit cell, as well as on values of $Q_1$.  



\section{Illustrative Examples}
\subsection{ Example 1: A Two-Phase Composite}\label{Exmpl1}

\begin{figure}[htp]
\centering
\includegraphics[scale=0.5, trim=3cm 5.75cm 3cm 4.5cm, clip=true]{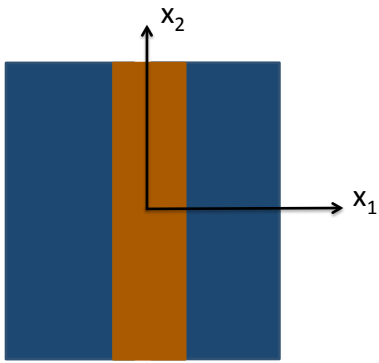}
\caption{The unit cell of a two-phase composite.}
\label{2-ph-Unit_Cell}
\end{figure}

We now examine the dynamic response of a \textit{two-phase} composite where the corresponding unit cell consists of a very stiff and a relatively soft layer; see Figure \ref{2-ph-Unit_Cell}.  
We show that, on the second frequency pass-band of such composites, the group and phase velocities in the $x_1$-direction (normal to layers) are antiparallel (\textit{backward wave}), whereas they are parallel in the $x_2$-direction (parallel to layers), signifying \textit{the negative-energy refraction with positive phase refraction} characteristic of this class of elastic composites in anti-plane shearing.   In contrast, on the first frequency pass-band of the composite, the group and phase velocities are parallel,  both in the $x_1$- and $x_2$-directions.

While the formulation and calculations are in terms of dimensionless quantities, in what follows the results are presented in terms of dimensional values  for the symmetric unit cell shown in Figure \ref{2-ph-Unit_Cell}.  
The dimensionless parameters and the results are calculated using the following specific material properties (typical for PMMA and steel):

\begin{enumerate}
\item $\mu_{1}=80\times 10^9$ Pa; $\rho_{1}=8000$ kg/m$^3$; total thickness = 1.3mm
\item $\mu_{2}=3\times 10^9$ Pa; $\rho_{2}=1180$ kg/m$^3$; total thickness = 3mm.
\end{enumerate}
The unit cell is 4.3 mm thick.  The resulting  dimensionless parameters used in the calculations have the following values:
\begin{enumerate}
\item $\bar{\mu}_{1}=3.0442$; $\bar{\rho}_{1}=2.4677$; $\bar{h}_2$ = 0.3023
\item $\bar{\mu}_{2}=0.1142$; $\bar{\rho}_{2}=0.3640$; $\bar{h}_1$ = 0.6977.
\end{enumerate}

The layers are isotropic, and subscripts $1, 2$  identify the properties ($\mu$ for shear modulus and $\rho$ for density) of each layer within the unit cell.  For example, $\mu_2$ stands for $\mu_{11}=\mu_{22}$ of layer 2. The superimposed bar denotes the corresponding normalized value; see equations (\ref{Average-1}, \ref{DimensionLess-1}). 
To obtain the frequency in kHz, and the group velocity, $v_{Jk}^g $,  in m/s, multiply $\nu$ by 105, and  $v_{Jk}^g $ by 2847, respectively.

%

\subsubsection{Frequency Band Structure}\label{Bands1}

Now examine the variation of the frequency as a function of the wave-vector components, $Q_1$ and $Q_2$. 
For each pair of $Q_1$ and $Q_2$, the direction of the wave vector is obtained from $\theta=atan(Q_2/Q_1)$,  the direction of the group-velocity vector from
$\alpha_J=atan(v_{J2}^g/v_{J1}^g)$ (which is shown below to be the same as that of the energy-flux vector,
$\beta_J=atan(E_{J2}/E_{J1})$), and the direction of the phase-velocity from 
$ \phi_J=atan(v_{J2}^p/v_{J1}^p)=\theta$,  respectively.  

\begin{figure}[htp]
\centerline{\includegraphics[scale=.5, trim=0cm 0cm 0cm 0cm, clip=true]{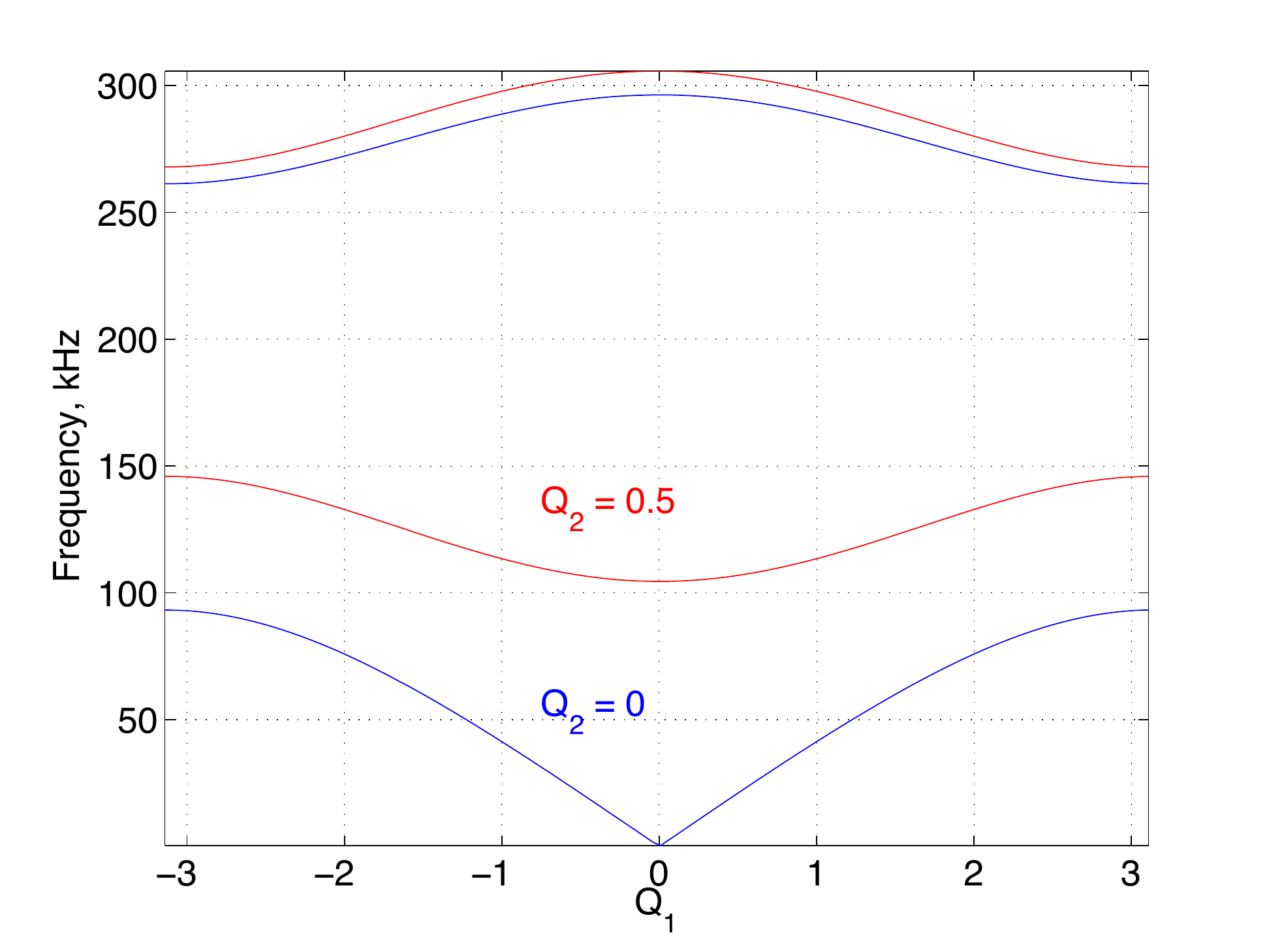}}
\caption{Frequency (in kHz) as function of $Q_1$ for indicated values of $Q_2$; two-phase composite.}\label{2ph-Q2-0-1}
\end{figure}
\begin{figure}[htp]
\centerline{\includegraphics[scale=.7, trim=0cm 2cm 0cm 2.9cm, clip=true]{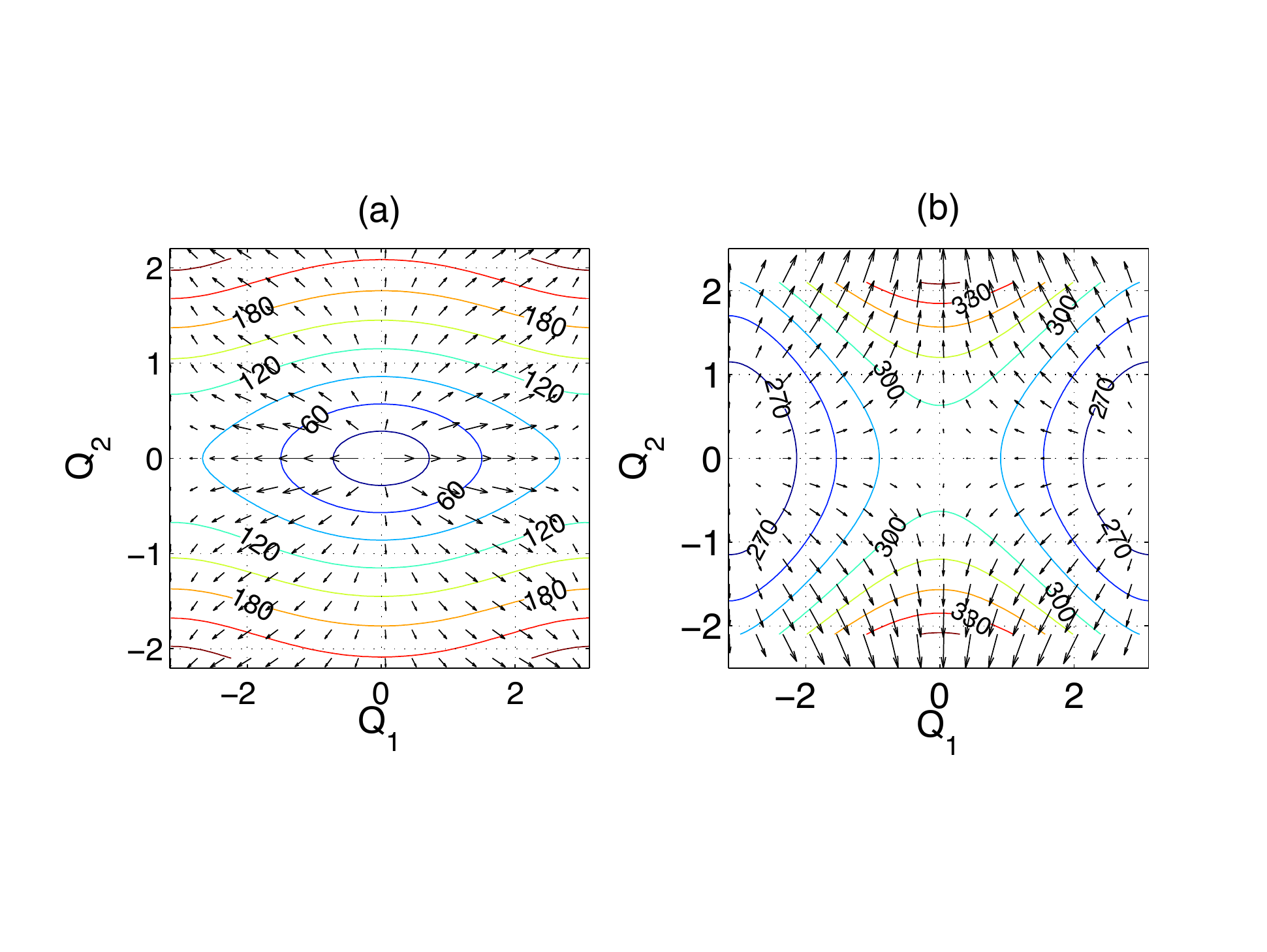}}
\caption{Contours of constant frequency (in kHz) and group-velocity vectors for: (a) first pass band, and (b) second pass band; two-phase composite; (for clarity, in graph (a) the $x_2$-component of the group velocity is reduced by a factor of 5).}\label{2ph-cont-GV-final-ab}
\end{figure}
Figure \ref{2ph-Q2-0-1} shows the frequency (in kHz)  as a function of $Q_1$ for indicated values of $Q_2$, and
Figures \ref{2ph-cont-GV-final-ab}(a, b) show the constant-frequency contours together with the direction of energy flow (superimposed arrows) as functions of $Q_1$ and $Q_2$, for the first two frequency pass-bands .  
As is seen,  for suitably small values of $Q_2$ these contours are ellipses on the first pass-band whereas they are  hyperbolae  on the second pass-band. 
On the first frequency pass-band, the corresponding components of the energy-flux and the phase-velocity vectors are parallel, but not on the second frequency pass-band.  
In this later case, the energy flux in the $x_1$-direction is antiparallel with the corresponding component of the
phase-velocity, while  in the $x_2$-direction these components are parallel. 
Hence the composite may or may not display \textit{negative refraction}, depending on the direction of the incident wave, as shown in 
Figures \ref{2ph-cont-GV-final-ab}$(b)$.
Here in addition, \textit{negative energy refraction is accompanied by positive phase refraction} and \textit{positive energy refraction is accompanied by negative phase refraction}; 
see subsection \ref{Exmpl2}  for illustration. 
In terms of the group and phase velocities, in Figure \ref{2ph-cont-GV-final-ab}(a),  
 $v_1^g v_1^p > 0 $ and $v_2^g v_2^p  >  0 $, whereas in Figure \ref{2ph-cont-GV-final-ab}(b),
$v_1^g v_1^p < 0 $ but $v_2^g v_2^p  >  0 $.
\begin{figure}[htp]
\centerline{\includegraphics[scale=.7, trim=0cm 2.5cm 0cm 3cm, clip=true]{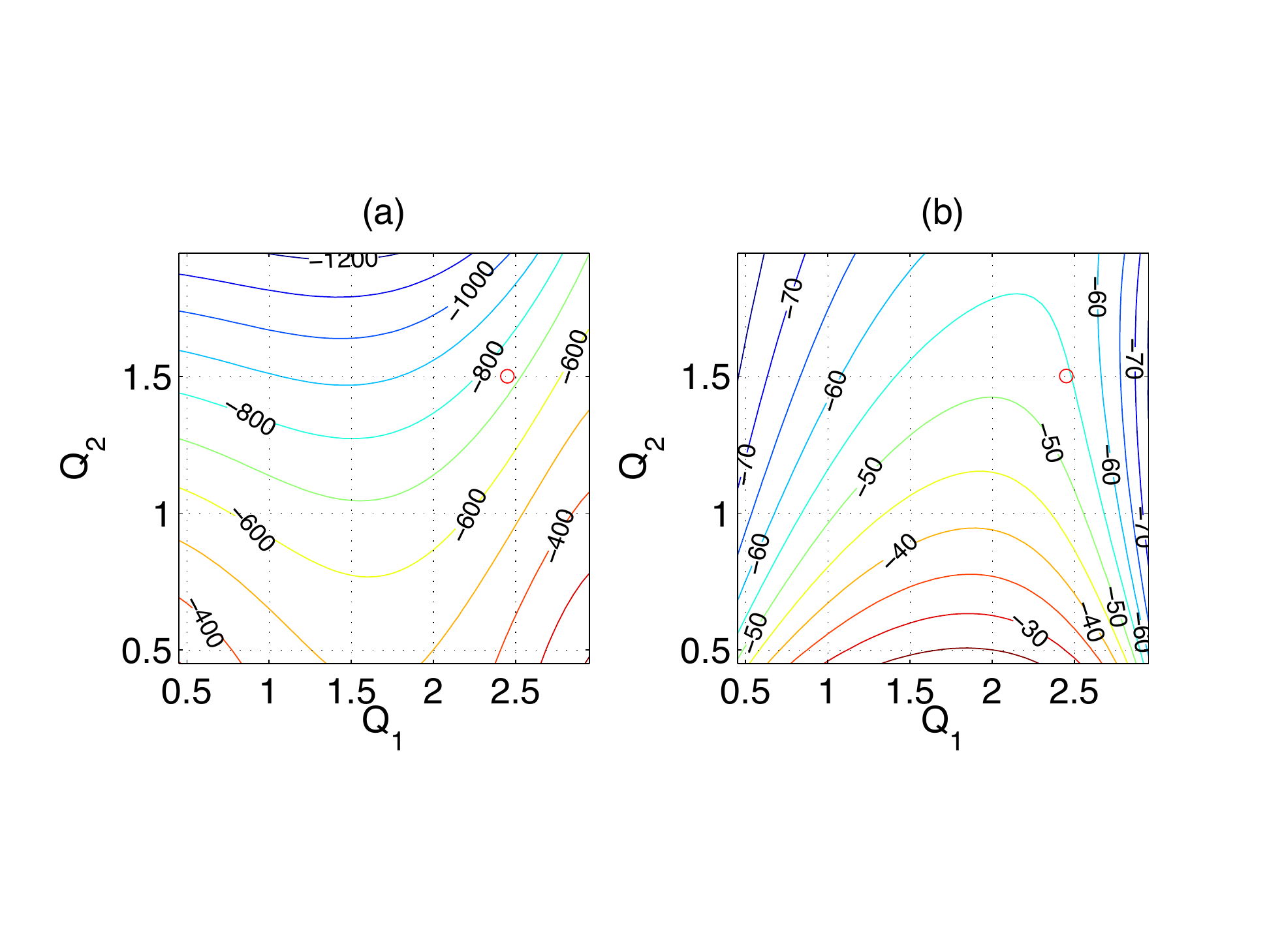}}
\caption{(a) Contours of constant group velocity (in m/s), and  (b) contours of constant (energy) refraction angle; for second pass-band of a two-phase composite (negative sign signifies  a negative angle with the $x_1$-axis). The red circles correspond to values for Example \ref{Exmpl2}.}\label{2ph-2b-vvg2-GVangle2-final-ab}
\end{figure}

Focusing on the second frequency pass-band, we have presented in  
Figure \ref{2ph-2b-vvg2-GVangle2-final-ab}(a) contours of constant  group velocity (in m/s) and in Figure (\ref{2ph-2b-vvg2-GVangle2-final-ab}b) those of constant (energy-flux) refraction angle.  As pointed out above, the group-velocity vectors are oriented in the direction of the energy flow.  
For positive wave-vector components, they have negative components in the $x_1$-direction but positive components in the $x_2$-direction; this is signified by negative signs in Figures \ref{2ph-2b-vvg2-GVangle2-final-ab}(a, b). 

\begin{figure}[htp]
\centerline{\includegraphics[scale=.5, trim=0cm 2.5cm 0cm 2.2cm, clip=true]{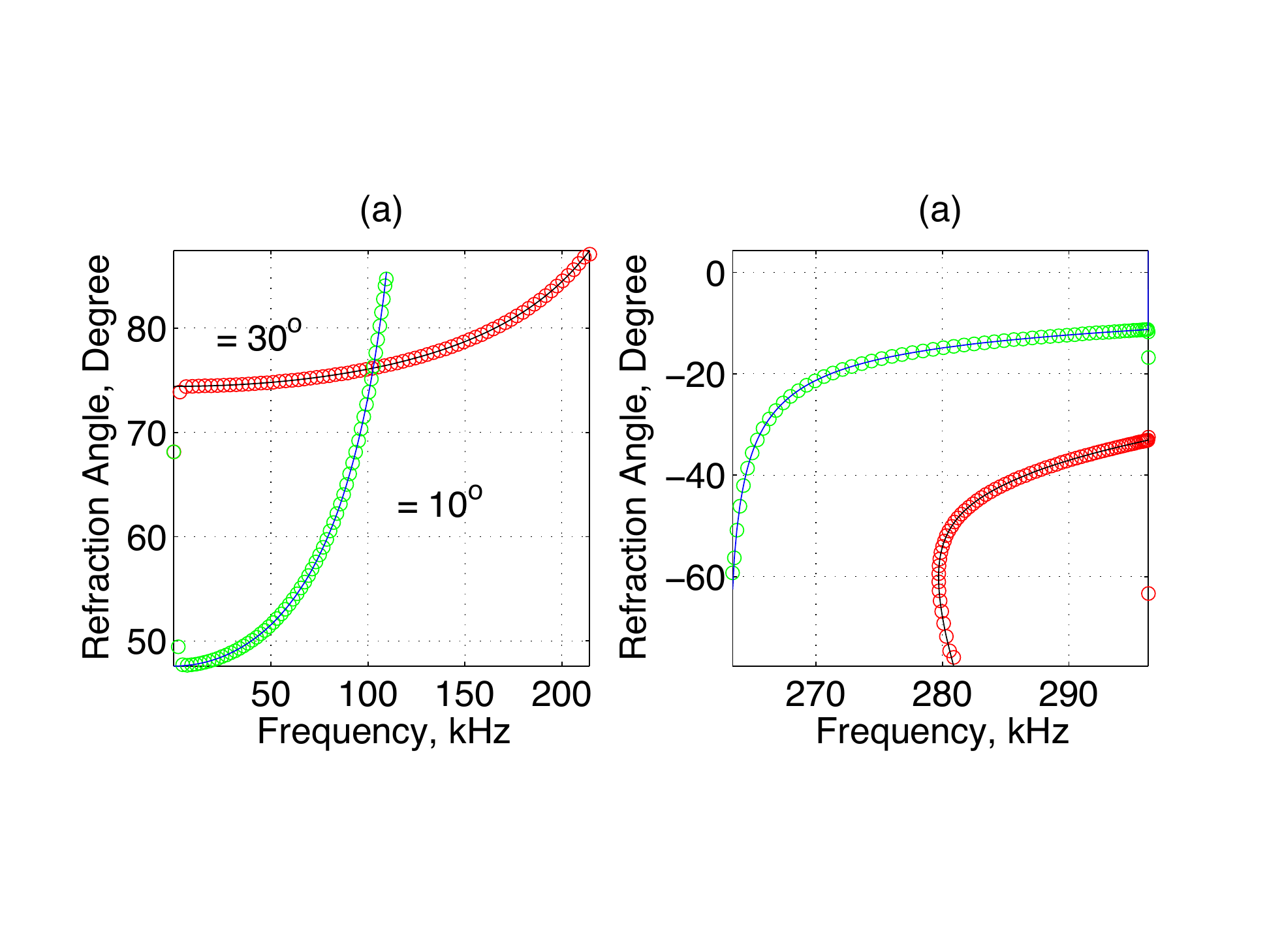}}
\caption{Refraction angle for  indicated incident angles as functions of frequency; solid (black: 30$^o$;  blue: 10$^o$) curves are energy-flux and open (red: 30$^o$;  green: 10$^o$) circles are for group-velocity direction; (a) first frequency pass-band, and (b) second frequency pass-band.}\label{2ph-vg-ef-angle-10-30-ab}
\end{figure}
As is mentioned before, the group velocity-vector defines the direction of energy flux.   Figures (\ref{2ph-vg-ef-angle-10-30-ab}a, b) show the refraction angles, ( $ \alpha_1$ and $\beta_1$) and ($  \alpha_2$ and $ \beta_2$),  as functions of the frequency for 
$ \theta=atan(\frac{Q_2}{Q_1})=10, 30^o $.   
The solid curves correspond to the energy-flux and the open circles to the group-velocity directions. The figures  show that the orientations of the group-velocity and energy-flux vectors are essentially indistinguishable.


\subsection{Example 2: Negative Refraction with Positive Phase Velocity and
Positive Refraction with Negative Phase Velocity 
}\label{Exmpl2}

\begin{figure}[htp]
\centerline{\includegraphics[scale=.6, trim=0cm 1cm 0cm 0cm, clip=true]{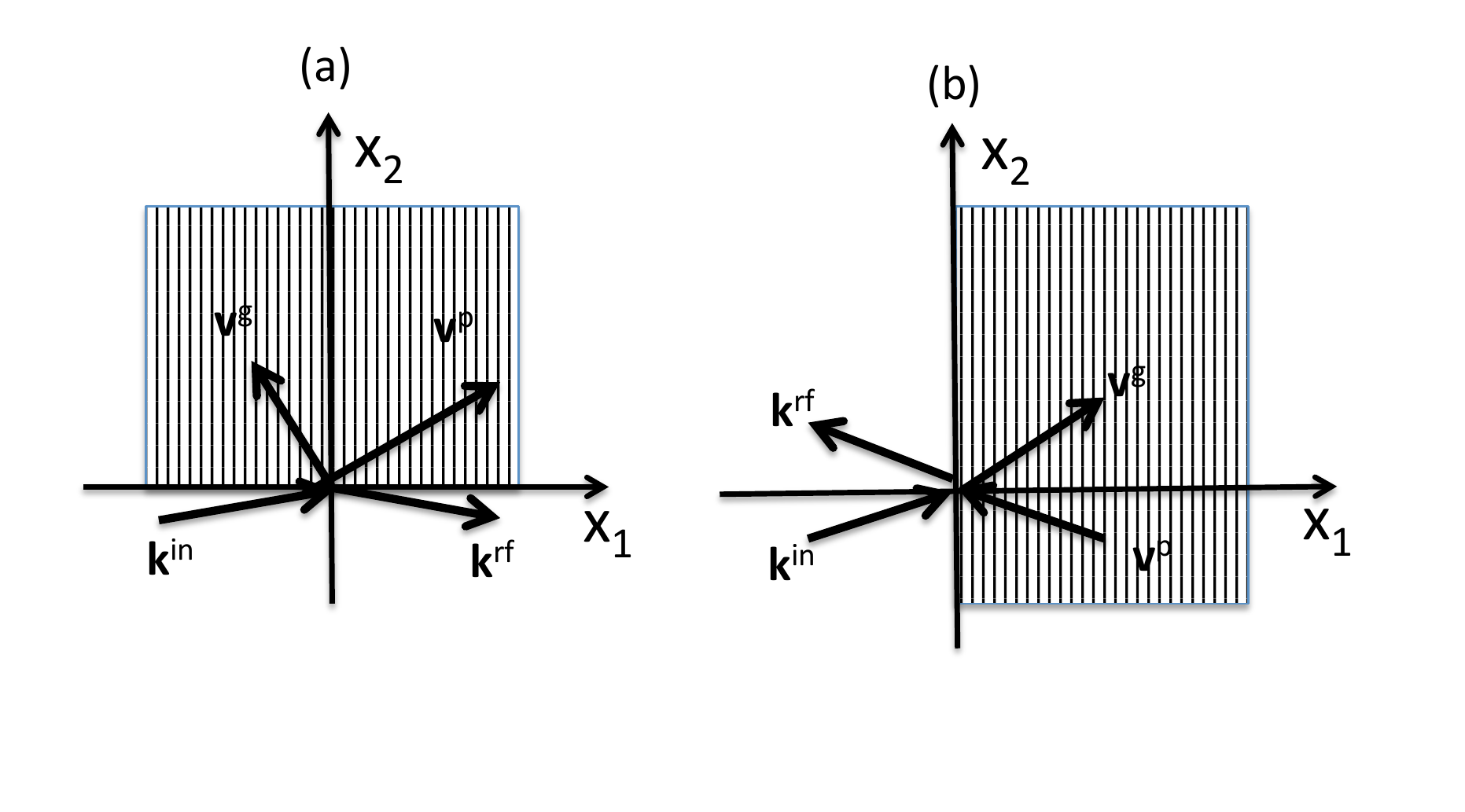}}
\caption{Plane harmonic wave of wave vector $\mathbf{k}^{in}$ is incident from
(a) $x_2 < 0 $ and (b) $ x_1< 0 $ homogeneous half-space toward a periodic half-space;  $\mathbf{k}^{rf}$  and  $\mathbf{k}^{tr}$  are  the reflected and transmitted wave vectors; $\mathbf{v}^{g}$ and $\mathbf{v}^{p}$ are the corresponding group and phase velocities.} \label{Negative_Positive_Refraction-ab}
\end{figure}

Examination of Figure \ref{2ph-cont-GV-final-ab} readily reveals that the layered composite can display negative refraction or negative phase-velocity refraction depending on how it is interfaced with a homogeneous solid. Figures  \ref{Negative_Positive_Refraction-ab}(a,b) suggest two possible ways. 
In Figure \ref{Negative_Positive_Refraction-ab}(a), the layered composite occupies the half-space $x_2>0$ while a homogeneous solid (say, aluminum) is occupying the half-space $x_2<0$, whereas in Figure  \ref{Negative_Positive_Refraction-ab}(b) the interface of the layered medium and aluminum is along the $x_2$-axis on the $x_1=0$-axis.  In each case, a plane harmonic anti-plane shear wave of wave-vector $\mathbf{k}^{in}$ is incident from the homogeneous solid toward the interface at an incident angle $\theta_0$, where $\theta=10^o$ in Figure  \ref{2ph-cont-GV-final-ab}(a) and $\theta=20^o$ in Figure  \ref{Negative_Positive_Refraction-ab}(b). In the first case, negative refraction is accompanied by positive phase refraction, and in the second case this is reversed, namely positive refraction is accompanied by negative phase refraction.

\begin{figure}[htp]
\centerline{\includegraphics[scale=.5, trim=0cm 0.2cm 0cm 0cm, clip=true]{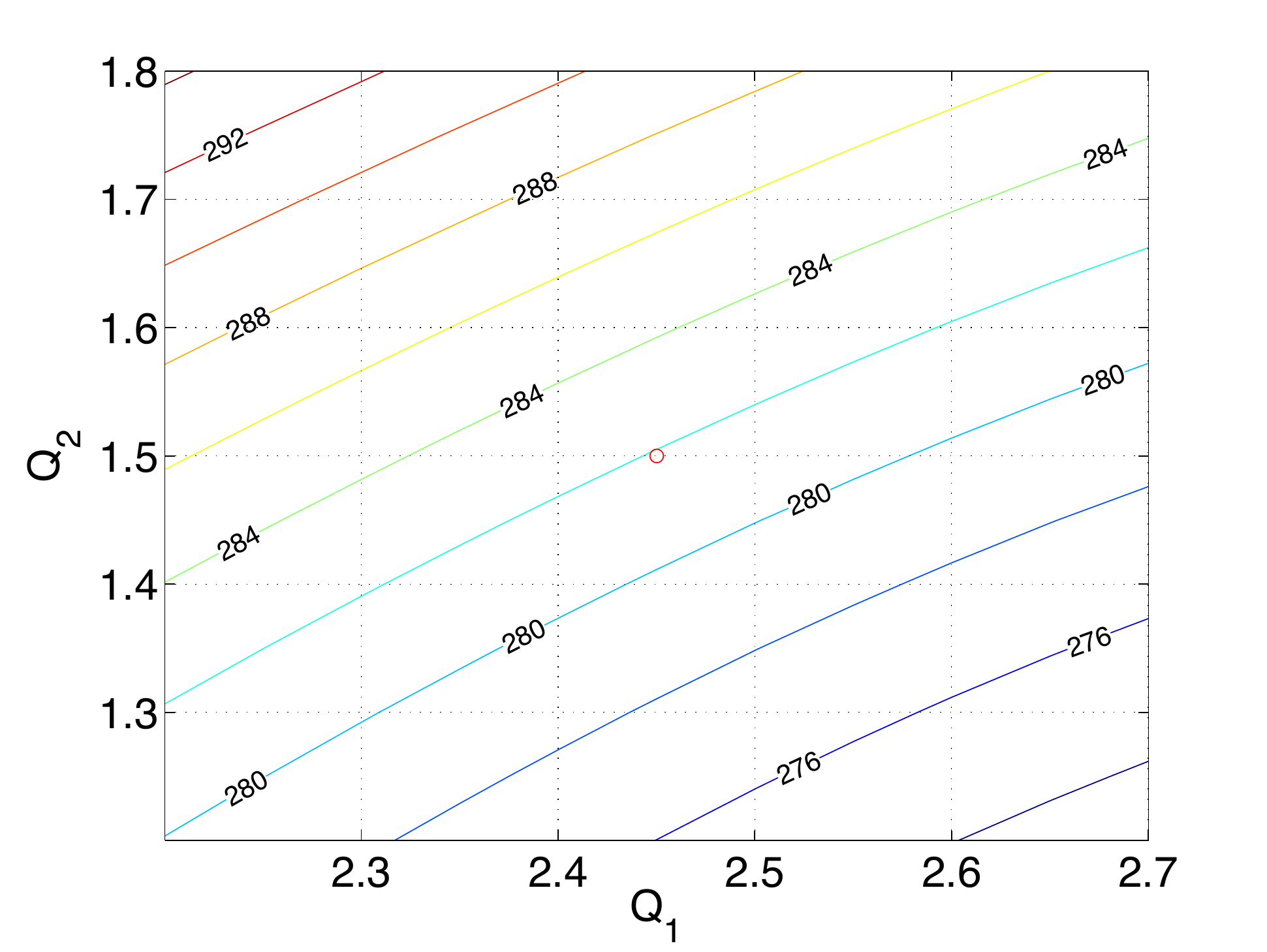}}
\caption{Contours of constant frequency (in kHz) for $2.4 < Q_1 < 2.6$ and $1.2 < Q_2 < 1.8$;  second frequency pass-band.}\label{2ph-cont-10degree}
\end{figure}

Let $C_0$ be the shear-wave velocity in the homogeneous solid. Consider 
Figure \ref{Negative_Positive_Refraction-ab}(a) and note that,
%
\begin{equation}\label{Phase-1}
Q_1=Q_1^{rf}=Q_1^{tr}=\frac{cos(\theta_0) }{\bar{C}_0}\nu,\quad
Q_2=-Q_2^{rf}=\frac{sin(\theta_0) }{\bar{C}_0}\nu,\quad
\bar{C}_0=C_0\sqrt(\bar{\rho}/\bar{\mu}_{11}),
\end{equation}
where $\bar{C}_0$ is the dimensionless value of the shear-wave speed in the homogeneous $x_2 < 0 $ half-space.   For a given frequency, say $\nu_0$, $Q_1^{tr}$ is given by (\ref{Phase-1})$_1$ and $Q_2^{tr}$ 
is calculated such that $\nu(Q_1^{tr},Q_2^{tr})=\nu_0$. 
For aluminum with a shear-wave speed of $C_0$ = 3,040 m/s and $\theta_0$ = 10$^o$,  Figure (\ref{2ph-cont-10degree}) shows the variation of the frequency in kHz for $2.2 < Q_1 < 2.7$ and $1.2 < Q_2 < 1.8$.
For a frequency $\cong$ 281.9 kHz ($\nu \cong$ 2.67), we have  $Q_1^{tr}=Q_1 \cong$ 2.47, and  $Q_2^{tr} =$ 1.50.  This gives a refraction angle of about -54.4$^o$ and a group velocity of $\cong$ 733.5 m/s. 
These values are identifyed in Figures \ref{2ph-2b-vvg2-GVangle2-final-ab} and 
\ref{2ph-cont-10degree} by red circles.

Now consider Figure  \ref{Negative_Positive_Refraction-ab}(b) and note that the phase angle $ Q_2\xi_2-\nu t $ on  the $ x_1 = 0 $-axis must be continuous.  Hence,
 \begin{equation}\label{Phase-2}
Q_2=Q_2^{rf}=Q_2^{tr}=\frac{sin(\theta_0) }{\bar{C}_0}\nu,\quad
Q_1=-Q_1^{rf}=\frac{cos(\theta_0) }{\bar{C}_0}\nu.
\end{equation}
For $\theta$ = 20$^o$ and a frequency $\cong$ 280 kHz ($\nu \cong$ 2.65), we have  $Q_2^{tr}=Q_2 \cong$ 0.85, and  $Q_1^{tr} =$ -1.50.  This gives a refraction angle of about 37.3$^o$ and a group velocity of $\cong$ 608.9 m/s. 

\subsection{ Example 4:  A Three-Phase Composite}\label{SUB3}

Qualitatively, the three-phase layered composite has the same dynamic response as the two-phase composite considered above. As an illustration, consider a unit cell consisting of a central 1 mm thick layer of steel that is sandwiched by two layers of polyurea/phenolic-microballoon composite of 0.4 mm thickness each, and then by two layers of PMMA of 1.25 mm each, the total thickness of the unit cell being 4.3 mm, the same as the two-phase composite. 

\begin{figure}[htp]
\centerline{\includegraphics[scale=.6, trim=0cm 0.5cm 0cm 0.4cm, clip=true]{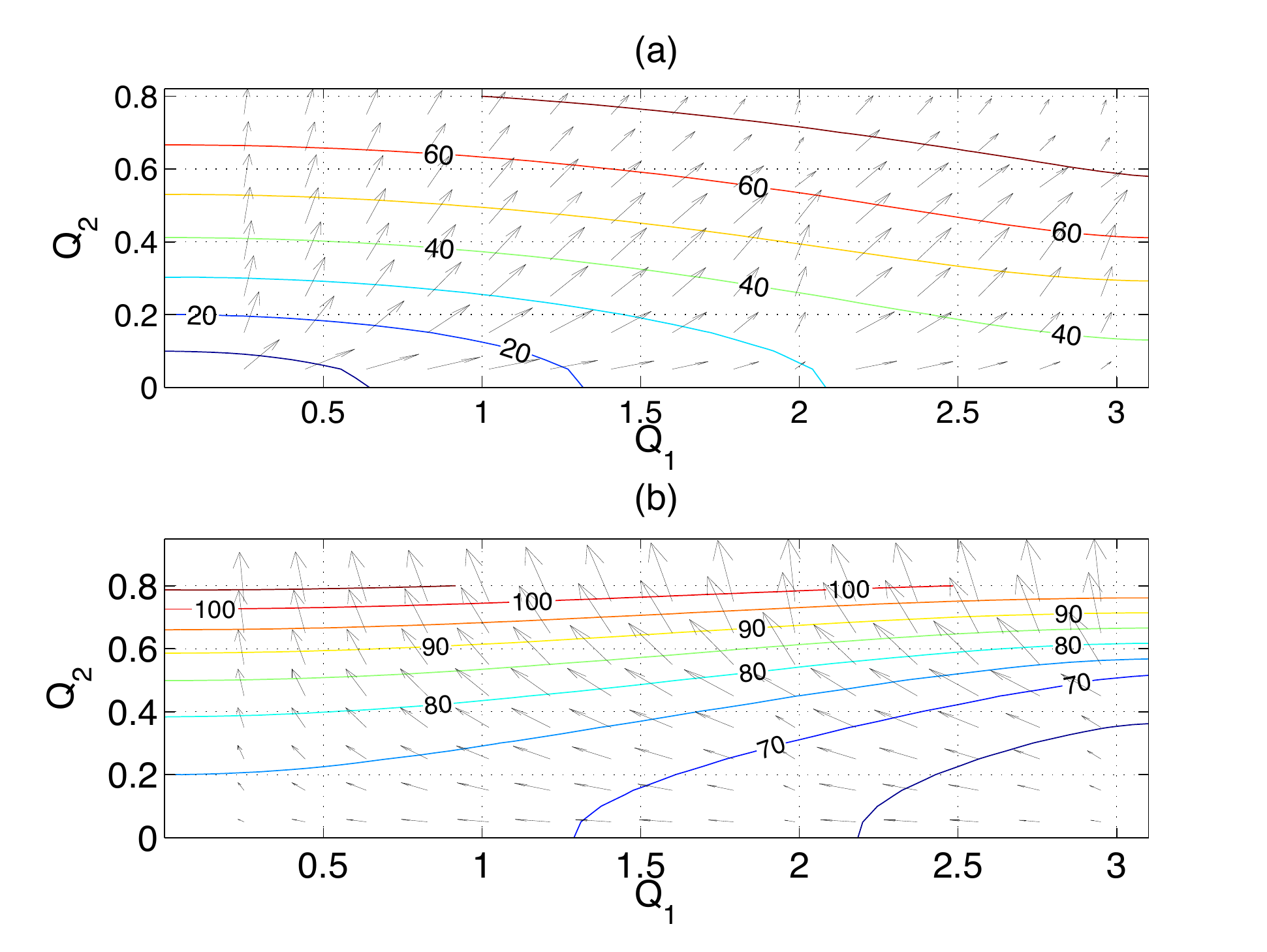}}
\caption{Contours of constant frequency (kHz): (a) first pass band, (b) second pass band, together with typical group-velocity vectors; the $x_2$-components of the group-velocity vectors are reduced by a factor of 10, and arrows are scaled for clarity;  (three-phase unit cell).}\label{3ph-cont-quiver}
\end{figure}

Figures (\ref{3ph-cont-quiver}a, b) display contours of constant frequency (in kHz) for the first (a) and the second (b) frequency pass-bands together with a representative number of group-velocity vectors; for clarity, the $x_2$-components of these vectors are reduced by a factor of 10.  For suitably small values of $Q_2$, these  contours are ellipses for the first pass-band, but they are hyperbolae for the second pass-band. 
Moreover, 
both of the  components of the group-velocity vectors are parallel with the corresponding phase-velocity components on the first pass-band, but on the second frequency pass-band, only their $x_2$-components are parallel while their $x_1$-components are antiparallel with the corresponding components of the phase-velocity vectors.  Hence, the refraction properties here are the same as those of the two-phase composites; see Figures (\ref{Negative_Positive_Refraction-ab}).  Also, the direction of the group-velocity and and energy-flux vectors are the same.

Hence, in general for anti-plane shear waves, layered periodic composites display \textit{negative refraction accompanied by positive phase-velocity refraction and  positive refraction accompanied by negative phase-velocity refraction} depending how they are interfaced with a homogeneous material.

\subsection{Discussion and Conclusions}\label{discussion}

Periodic elastic composites can be designed to have static and dynamic characteristics that are not shared by their constituent materials.   
Some of the dynamic characteristics and responses of layered periodic composites are explored in this work, using harmonic anti-plane shear waves.  The considered composites lack periodicity in the direction parallel to the layers.  This profoundly affects their dynamic response, leading to anomalous wave-refraction,  namely, for this class of composites, negative refraction is accompanied by positive phase-velocity refraction, a phenomenon which was first recognized by \cite{meisels2005refraction}.
We have in addition shown that the composite can also display negative phase-velocity refraction accompanied by positive energy refraction, a phenomenon which does not seem to have been recognized before.
  
In the present work, a general variational approach is developed that produces the entire band structure of the composite for unit cells of any number of layers with any arbitrary properties.  
Explicit expressions are developed for the
band structure, group-velocity and energy-flux vectors.
The general results are illustrated using a two-phase and a three-phase unit cell with piecewise constant properties.  The presented method is applicable and effective also when some or all of the layers in a unit cell have spatially varying properties. 
 
\textbf{Acknowledgments}:
This research has been conducted at the Center of Excellence for Advanced Materials (CEAM) at the University of California, San Diego, under DARPA AFOSR Grants FA9550-09-1-0709 and RDECOM W91CRB-10-1-0006 to the University of California, San Diego.

\section{References}

\bibliographystyle{elsarticle-harv}
\bibliography{REFS_Harmonic-1}
%

\end{document}